# Longitudinal Trends in Networks of University-Industry-Government Relations in South Korea: The Role of Programmatic Incentives

Han Woo PARK

Dept. of Media & Communication, YeungNam University, South Korea

hanpark@ynu.ac.kr ; http://www.hanpark.net; http://english-webometrics.yu.ac.kr

Loet Leydesdorff

Amsterdam School of Communications Research, University of Amsterdam,

Kloveniersburgwal 48, 1012 CX  Amsterdam, The Netherlands

loet@leydesdorff.net; http://www.leydesdorff.net

*Abstract*

This study examines the longitudinal trend of systemness in networked research relations in South Korea using a triple-helix (TH) indicator of university-industry-government (UIG) relations. The data were harvested from the *Science Citation Index* (SCI) and its counterparts in the social sciences (SSCI) and the arts and humanities (A&HCI). The total number of Korean SCI publications has grown rapidly since 1965. However, the TH indicator shows that the network dynamics have varied considerably according to the research policies of the national government.  The collaboration patterns, as measured by co-authorship relations in the SCI noticeably increased, with some variation, from the mid-1970s to the mid-1990s. However, inter-institutional collaboration in the first decade of the 21$^{st}$ century was negatively influenced by the new national science and technology (S&T) research policies that evaluated domestic scientists and research groups based on their international publication numbers rather than on the level of cooperation among academic, private, and public domains. The results reveal that Korea has failed to boost its national research capacity by neglecting the network effects in science, technology, and industry.

**Introduction**

The origins of the concept of national systems of innovation in East Asia can be traced to Freeman (1987, 1988), who drew "lessons from Japan" after visiting that country. Lundvall (1988) generalized the concept of national systems of innovation by proposing a coordination system based on *interactions* between users and producers. He said that "it is obvious that any model of a national system of innovations must take into account the interaction between universities and industry" (p. 364), while "government may intervene, directly or indirectly, in relation to the establishment and restructuring patterns of user-producer relations" (p. 358). The model of a triple helix (TH) of university-industry-government (UIG) relations (Etzkowitz & Leydesdorff, 2000) enables us to study both the bilateral and trilateral interactions among these three institutional spheres in an innovation system.

More recently, the availability of a TH indicator—namely, the mutual information among uncertainties in three dimensions—enables us to study the extent to which networks of relations among the three spheres have developed into a synergetic configuration. This methodology can also be used for the analysis of innovation systems other than nationally defined ones, such as sectoral, technological, and regional innovation systems (Carlsson, 2006). In other words, the indicator enables us to study empirically the question of whether and to what extent various possible systems are integrated and/or differentiated.

In this paper, we focus on a specific national system: South Korea. Because of its history, South Korea can be expected to entertain a *national* system of innovations. The country is still struggling with a tenuous past in relations with its neighbors. However, in recent decades both universities and industries have developed not only nationally, but also internationally. Subsequent national governments have undertaken a variety of interventions in order to strengthen the knowledge base of innovations in terms of university-industry relations and to stimulate the development of different sectors. More recently, these stimulation programs have been extended to the social sciences and the humanities.

The Korean government is acutely aware of the need to bolster the Korean national system of innovations. This builds on a tradition of attention to science and technology (S&T) policies which goes back to the dictatorship of the 1970s, but is reinforced by the relative isolation of the country among the Organisation for Economic Co-operation and Development (OECD) nations. Unlike European Union (EU) member states, Korea is not part of a similar supra-national arrangement. Whereas Japan could traditionally rely on more horizontal mechanisms of integration (Irvine & Martin, 1984; Yamauchi, 1987), the Korean government's policies are important for shaping and continuously reconstructing the system.

For example, in response to structural problems coordinating the national research and development (R&D) administration, President Lee Myung-Bak (2008-present) formulated a grand strategy for reforming the Korean R&D system by reorganizing governmental agencies: the new government launched the Ministry of Education, Science and Technology (MEST) and the Ministry of Knowledge Economy (MKE). While MEST is intended to play a role as key coordinator for incorporating the human resources of education, science and technology for the purpose of national development, MKE has a mission in fostering knowledge-based innovation capacities across the country (Choi, 2008; Lee, 2009a).

The effectiveness of government intervention, however, cannot be taken for granted. The systems can be expected to have resilience in following their own institutional logic in the case of academia and to be driven by market forces in the case of industry. In a recent study of university-industry co-authorship relations in Japan, Leydesdorff & Sun (2009) found a long-term erosion of university-industry co-authorship despite a series of government programs directed at their stimulation. In this study, we follow the design of the Leydesdorff & Sun study by operationalizing relationships in terms of co-authorships in the formal SCI literature. This data enables us to construct time-series with three relevant dimensions—university, industry, and government—as institutional addresses that we can analyze in terms of bilateral and trilateral relations. Although co-authorship relations in the formal literature only show the tips of icebergs of possible exchange relations underneath, they have widely been used and validated as a proxy for the dynamics of these relationships (Wagner, 2008).

Surowiecki (2005) postulated three conditions for successful scientific collaboration: diversity, independence, and decentralization. But how can one measure the degree of 'independent but competing, and mutual information transmission' embedded in the research collaboration occurring in a scientific community? In our opinion, the TH indicator of UIG relations is particularly useful for examining how effectively individual actors in science work together across institutional boundaries and the consequential status of the interaction-based knowledge infrastructure. More specifically, as shown in previous research (Etzkowitz & Brisolla, 1999; Leydesdorff, 2003; Leydesdorff & Fritsch, 2006; Leydesdorff *et al.*, 2006; Park *et al.*, 2005, Shapiro, 2007), the TH model of the three independent UIG actors has the capability to capture both the dynamics within the three helices and new developments at the network level, generated and stabilized in mutual information exchanges among the helices.

The research question of this study addresses the relation between government policies and the development of networked systems of relations among the TH partners from a longitudinal perspective. How often do researchers working in different institutions in Korea collaborate with each other? What pattern do their co-authorship relations reveal? Is this pattern a result of social (e.g., institutional), economic or political factors? To what extent and how have government policies affected these patterns? Several studies have investigated the TH models of Western countries, but only a few have examined Asian countries in terms of co-authorships in the international literature (Leydesdorff & Sun, 2009; Park *et al.,* 2005; Shapiro, 2007). In recognition of the sustained and remarkable growth of the research output in Asian countries in terms of the share of publications and citations in the SCI database (Leydesdorff & Zhou, 2005), we examine Korea's national research portfolio from this TH perspective. Based on the results, some specific characteristics of the Korean national research system can further be discussed.

**Theoretical Framework and Relevant Literature**

The structure of a national R&D system can be examined through the use of a TH framework that studies the relations among academia, business, and government (Etzkowitz, 2008). TH network analysts argue that UIG interactions represent the core of knowledge-based innovation with circulation among and within the three spheres. Bilateral and trilateral relations can stimulate ideas and policies across institutional boundaries. In the theme paper at the 2009 TH conference (Glasgow, June 2009), Dzisah and Etzkowitz (2009) emphasized that the TH of UIG joint projects makes it possible to stimulate the knowledge-based strategy and speed the rate of socioeconomic development by enhancing the free flow of people, ideas and innovations in the national S&T capacity of R&D systems. In other words, facilitating these dynamic relationships and interplay may be a first step in creating the necessary and sufficient conditions for further innovation and sustainable development in a national or regional system.

Despite the accumulation of substantial knowledge and experience with UIG interactions in Korea, Shapiro (2007) noted that "the TH paradigm has not been extensively applied to the Korean case, as persuasive studies of R&D in Korea have traditionally been conducted in terms of the National Innovation System approach" (p. 171). A review of the recent literature (since 2000) provided only a few examples of studies on the development of UIG relations in Korea. For example, in their cross-national TH study of Korea and Brazil, Etzkowitz and Brisolla (1999) described the adoption by these two countries of similar research policies in order to rapidly upgrade their national innovation systems. Both governments used technology-bolstering actions and public interventions in the development and diffusion of new technologies. Furthermore, TH relations were arranged in both countries on the basis of exchanges among UIG entities. The exchange materials included various forms, such as resources, intangible information, interpersonal interaction, and visible goods.

Particularly, a TH structure in the R&D field can be considered as a set of collaborative relations among UIG entities in terms of research practices (e.g., authorship, citation). In a similar vein, Park *et al*. (2005) compared the network-based innovation systems of Korea and the Netherlands using three TH indicators: scientometric, technometric, and webometric measures. These authors concluded that, "Despite the increasing amount of scientific and technological outputs in terms of the knowledge-based dynamics, South Korea's portfolio is more traditional than that of the Netherlands in both the public and private sectors" (p. 25).

On the basis of his studies of the Korean TH network, Shapiro (2007) suggested that Korea's policy-makers could develop programs to facilitate the cooperation between academia, industry, and government in various TH dimensions and build new forms of social capital into the national research system. A recent article by Trotter *et al*. (2008) showed that social capital, such as communication, trust, and conflict, plays an important role in the health of collaborative partnerships among university research centers, private-sector firms, and other strategic institutions. Shapiro (2007), however, did not provide detailed information about the differentiation and integration among the three UIG institutions. In contrast, our research offers a comprehensive TH analysis in terms of long-term data collection and detailed classification. In spite of the increasing importance of China and South Korea in science and technology, the national R&D portfolios of these countries have not yet been analyzed in terms of their TH dynamics.

**Triple Helix Development and South Korea**

As nations compete against one another to stay ahead in the game of scientific and technological innovation and knowledge-based developments, conducting world-class research and publishing its results become increasingly important. The goal of our study is to identify the underlying knowledge-linkage patterns of UIG relations in Korea's national research system. As reported in a recent publication (Park & Leydesdorff, 2008), Korea rapidly increased its share of SCI-listed publications. In terms of the number of papers in the SCI-listed journals, Korea occupied the $12^{th}$ position in 2007. This is an impressive jump from $21^{st}$ place in 1996. However, only three scientists working in Korean institutions were included in the 2008 list of the Institute for Scientific Information's (ISI) database of *Highly Cited Researchers*. The relatively low numbers of citations obtained by Korean scientists is largely due to a national research policy that emphasizes publication numbers for individual researchers and deemphasizes the incorporation of a wide variety of informal ideas (e.g., consulting) and industrial resources (e.g., patents) across institutional boundaries (Hwang *et al*., 2004; Kim, 2008; Kwon, 2009a; Kim & Nelson, 2000; Lee, 2009b; Lee, 2000; Park & Leydesdorff, 2008).

For academics, the importance of effective collaboration with researchers in the governmental and commercial sectors is growing in science, engineering, and the social science fields, largely due to the researchers' access to tremendous amounts of data as well as extensive in-house research capacities (Coleman, 2007). Mahmood and Singh (2003)—in their analysis of the innovative capability of six Asian countries, including Korea, Taiwan, Singapore, India, China, and Hong Kong—found that Korea successfully managed to shift from an economy based on 'heavy and chemical industries' to one based on 'knowledge and technology'-intensive industries over the last three decades. Nevertheless, there are still gaps in the national system of innovations across sectors. Korea's commercial sector is very active in contributing innovation to the national research system through patents, but it does not share much of its R&D capabilities with academic or governmental entities. In other words, Korean society's R&D activities, including industrial patents and scientific publications, are not well integrated at the national level.

The need for such integration is not restricted to the natural sciences. Increasingly, public and business agencies have resources to gather information on many subjects, such as electoral and market surveys, land registries, crime statistics, and consumer behavior profiles. Academic social scientists have been encouraged to collaborate with governmental and industrial research in order to employ the available public and commercial data in developing sophisticated theories of high quality. Savage and Burrows (2007, p. 886) observed, "Fifty years ago, academic social scientists were seen as occupying the apex of the—generally limited—social science research 'apparatus'. Now they occupy an increasingly marginal position in the huge research infrastructure that forms an integral feature of what Thrift (2005) characterizes as a *knowing capitalism* where circuits of information proliferate and are embedded in numerous kinds of information technologies." These authors concluded that social scientists have tended to err in emphasizing their abilities to be more reflexive and sophisticated than their governmental and industrial colleagues. In a similar vein, Moon and Lee (2005) suggested the combined use of three R&D input and output elements (academic sector, public institute, industry) according to two separate fields (natural science and social science) when comparing Korea's S&T capabilities with those of five developed countries (France, Germany, Japan, UK, US).

The Korean research system seeks to increase R&D activities to narrow the gap with Western countries in the short term, while the research strategy of Western nations is oriented towards increasing their accumulative capacity and quality of R&D over the long term (Lee, 2000; Mahlich & Pascha, 2007; Wade, 1990). Given Korean incentives, such indices as the number of scientific publications and patent registrations show a surprising increase for Korean scientists, but the production of high-quality outputs in S&T still lags behind that of advanced nations (Choi, 2008; Lee, 2009a; Lee & Kwun, 2003). Some analysts have argued that Korea's national R&D programs are disjointed because they are funded by several different agencies with overlapping policy targets (Hong, 2005; Hwang, 2002; Song *et al*., 2007). Although there is a Korean National S&T Council (NSTC), its objectives are confined to the natural and engineering sciences. The Korean government does not have a strong steering organization for coordinating R&D programs across institutional and disciplinary borders.

We mentioned previously that in response to these structural problems in coordinating the national R&D administration, President Lee Myung-Bak (2008-present) formulated a grand strategy for reforming the Korean research system by organizing two new ministries. The R&D policies of the current government, however, are beyond the scope of this research. We focus on the development of the policy environments from the Park Jung-Hee to the Roh Moo-Hyun governments (1970-2007) and investigate the longitudinal changes in collaborative research activities among Korean universities, governmental institutions, and commercial R&D organizations based on interactions with institutional and policy settings.

## Brief History of Governmental R&D Policies

This section briefly introduces the historical development of the Korean R&D system broken down into government periods from 1970s to the early 2000s. UIG relations can be considered as a knowledge infrastructure where three institutional actors interact as relatively independent entities. However, as Dzisah and Etzkowitz (2009) noted, the coordinating role of government in both developing and developed societies is key to improving the conditions for active collaboration among institutional spheres. Kwon (2009b) acknowledged that the Korean government has taken strong initiatives to direct national R&D activities. Under these circumstances, the responses of universities, public institutes, and industry can be expected to vary with different government policies, but in relation to evolving stages of the Korean R&D systems.

**Table 1**. Characteristics of R&D programs according to government policies.

| Government | Characteristics of R&D programs related to the TH indicators |
|---|---|
| Park, Jung-Hee (1970-1979) | Government's strong push to run governmental institutes and joint research between universities and public organizations |
| Chun, Doo-Hwan (1980-1987) | Restructuring of government-sponsored research institutes; e.g., the integration of the KAIS (Korea Advanced Institute of Science) university and the KIST (Korea Institute of Science and Technology) into the KAIST (Korea Advanced Institute of Science and Technology) |
| Roh, Tae-Woo (1988-1992) | The gradual opening of research organizations in both private and public sectors; e.g., KIST became independent from KAIST in 1989 |
| Kim, Young-Sam (1993-1997) | Dominance of governmental agencies from early 1990 to 1997 when Korea started to be subject to the conditions of the International Monetary Fund (IMF) |
| Kim, Dae-Jung (1998-2002) | BK21 project started in 1999 to increase the research capacity of universities through large central government subsidies, thus decreasing UIG joint research |
| Roh, Moo-Hyun (2003-2007) | Continual promotion of the BK21 and internationalization of R&D, particularly in the academic sector. PBS was introduced in the governmental sector. |

Source: Hong (2005), Hwang (2002), Kwon (2009b), and other sources.

Some of the unique characteristics of Korean R&D programs by government are summarized in Table 1. President Park, Jung-Hee, who can be called both 'the father of modernization' and a 'military dictator' (Oberdorfer, 2002), was president during the 1970s. He actively established a national S&T infrastructure. This president strongly believed that advancement of the national economy could not be achieved without the development of an indigenous S&T research capacity. The Ministry of Science and Technology and the 16 public research institutions, including the Korean Institute of Science and Technology (KIST), had been founded in the 1960s. The mutual cooperation between universities and governmental researchers was coordinated by the policies of the 1970s with the goal of elevating Korea to the level of a developed country.

Compared to this prior government, President Chun, Do-Hwan (1980-1987), is probably underestimated. He emphasized the unique role of public R&D organizations in strengthening the national research system. Government-sponsored research institutes were reorganized during this period (Kim, 2006). During his term, industrial researchers no longer worked in isolation, but started to collaborate with academic colleagues, as previously strict regulations of the university system became weaker. Furthermore, the world famous Steel Company POSCO established a research-intensive university POSTECH (Pohang Institute of Science and Technology) to strengthen its R&D activities. From this perspective, one can question whether more relaxed regulation of university research by the government might have stimulated university-industry (UI) relations relative to the previous period. We will return to this issue in the results section.

A new environment for university-government cooperation was again created during the next administration of President Roh, Tae-Woo (1988-1992). During this period, national R&D programs were diversified. For example, the former KIST became independent from KAIST. Furthermore, 'the master plan for advancement of basic research in S&T' drafted in 1989 emphasized fostering mutual relations between public and university laboratories.

President Kim, Young-Sam (1993-1997) continued the policy of his predecessor; he also reinforced national S&T projects and diversified government R&D programs. The stimulation of UIG collaboration remained an important policy objective, similar to that of the previous government. The Kim administration responded to the transition from the industrial age to the information society through industrial restructuring and technological innovation programs. For example, research-intensive universities were founded: GIST (Gwangju Institute of Science and Technology) in 1993, KIAS (Korea Institute for Advanced Study) in 1996, and ICU (Information and Communications University) in 1998.

In response to the financial crisis in Korea at the end of 1997, Korean companies reduced their R&D investments. The government of President Kim, Dae-Jung (1998-2002) accordingly began to establish policy measures stimulating the research activities of university academics. The 'Brain Korea 21' project (hereafter, BK21) can be considered as the main government incentive during his term. BK21 was launched mainly to encourage university researchers, particularly young faculty members and postgraduate students, to produce high-quality research output that could be published in internationally peer-reviewed journals (Moon & Kim, 2001; RAND, 2007). Indeed, the research profile of Korean university academics was boosted in terms of the number of papers in the journals covered by SCI. According to an official website about the achievement of BK21 (http://bnc.krf.or.kr/home/eng/bk21/achievement.jsp), the number of SCI-listed papers published by BK21-sponsored researchers increased from 3,765 in 1998 to 7,281 in 2005.

Stimulated by the soaring performance of university academics in terms of the numbers of papers in journal covered by the SCI, the government of President Roh, Moo-Hyun (2003-2007) further promoted the BK21 project in order to empower young scientists and postgraduates to participate in the internationalization of R&D. During this period, a so-called Project-Based System (PBS) was introduced at public research facilities. Under this scheme, researchers in public institutes were evaluated based on performance, such as income from contracts. The purpose of this policy was to stimulate governmental research institutes to focus on conducting highly qualified research that leads to convergent, complex, and sophisticated technology development, in order to raise the knowledge intensity of Korean industries. Another characteristic of this period is that university-industry cooperation became increasingly oriented towards the education of industrial engineers and the production of innovative patents. This policy went under the name of 'stimulating entrepreneurial universities.'

In summary, the subsequent Korean R&D programs can be characterized as follows. First, there has been continuous government involvement in restructuring the national research system by allocating funding and modifying evaluation indicators. Second, there have been considerable changes in the main role of the universities. According to Kwon (2009b), until the early 1990s, the Korean universities were regarded as human resource trainers to meet industrial needs rather than as important research capacities in the national innovation system. For the revitalization of university research, various R&D programs have been implemented by Korean governments since the early 1990s and consequently the increase of university outputs in terms of international papers has been impressive.

This rapid growth of the academic sector may have constrained TH developments at the national system level by generating more differentiation than can be managed in collaborations among the three TH actors. Moreover, as some scholars (Hong, 2005; Hwang, 2002; Kwon, 2009b) have pointed out, since the mid-1990s, the funding principle of 'selection and concentration' under which a winner-take-all principle prevails may have impoverished UIG collaboration. In other words, we expect that the academic, private, and public sectors are more likely not to be sufficiently integrated at the network level to produce a R&D communication system within Korea. Our results can be considered as confirming this hypothesis.

## Methods and Materials

The data were collected using the Web of Science provided by the ISI of Thomson-Reuters. All papers in the SCI, *Social Science Citation Index* (SSCI), and the *Arts and Humanities Citation Index* (A&HCI) with at least one Korean address were collected. All Korean addresses were manually attributed to the three UIG institutions. Note that some institutions in South Korea can be classified as both university and government. While research-oriented universities such as KAIST and GIST were labeled as academic, the KIST was labeled as government based on its functions.[1]

For the analysis of this data, we used entropy statistics. When variation is considered as a relative frequency or probability distribution ($\Sigma_i\, p_i$), the Shannon-type information or the uncertainty contained in the distribution (*H*) is defined (Shannon, 1948; Shannon & Weaver, 1949) as follows:

$$H_i = -\Sigma_i\, p_i \log_2 (p_i) \tag{1}$$

Equivalently, for a two-dimensional distribution $H_{ij}$ is:

$$H_{ij} = -\Sigma_i \Sigma_j\, p_{ij} \log_2 (p_{ij}) \tag{2}$$

---

[1] The data also contains a category "Others" which will not be used in our analysis. Some unidentified institutions (that cannot be classified into the three UIG categories) were also classified under this category.

This uncertainty is the sum of the uncertainty in the two dimensions diminished with their mutual information. In other words, the two variations overlap in their co-variation and condition each other in the remaining variations.

Mutual information can be written in information theory using the *T* of transmission between two distributions as follows:

$$H_{ij} = H_i + H_j - T_{ij} \qquad (3)$$

$$T_{ij} = H_i + H_j - H_{ij} \qquad (4)$$

$T_{ij}$ is zero if the two distributions are completely independent (i.e., the co-variation is zero), but otherwise necessarily positive (Theil, 1972, pp. 59f.). Abramson (1963, p. 129) derived that the mutual information in three dimensions—let us use "u" for university, "i" for industry, and "g" for government—can be defined analogously as follows:

$$T_{uig} = H_u + H_i + H_g - H_{ui} - H_{ug} - H_{ig} + H_{uig} \qquad (5)$$

The resulting indicator can be negative or positive (or zero) depending on the relative sizes of the contributing terms. A negative value means that the uncertainty prevailing at the network level is reduced. McGill (1954) proposed calling this possible reduction of uncertainty "configurational information" (Jakulin & Bratko, 2004; Yeung, 2008).[2] Because the information is configurational, the reduction of uncertainty cannot be attributed to one of the contributors. These network effects are systemic. Note that the bilateral terms contribute to the reduction of uncertainty, while uncertainty in three dimensions adds to the uncertainty which prevails at the network level.

---

[2] Yeung (2008, pp. 51 ff.) formalized this measure as $\mu^*$.

For example, it matters for a government whether or not universities and industries already entertain strong relations. Similarly in a family system, it matters for a child whether the relationship between the two parents is in good shape. When structure is provided by the relationship between two of the three partners, uncertainty is reduced for the third. For example, knowing the answer of the one partner on a question, can then inform us about the likely answer of the other. This possible reduction of uncertainty in a TH configuration can be measured using the mutual information in three (or more) dimensions.

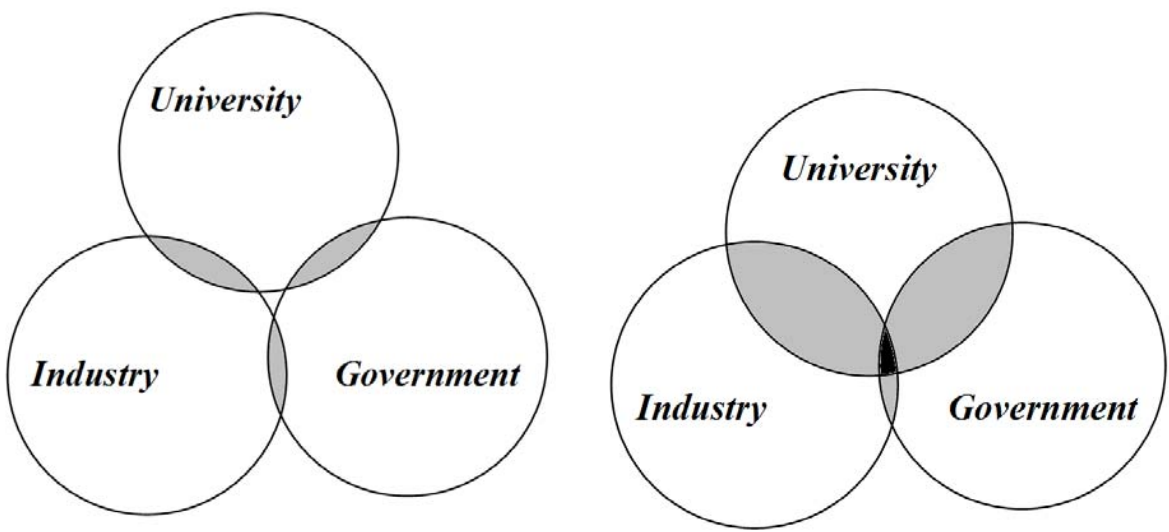

**Figure 1**: A TH configuration with negative and positive overlap among the three subsystems.

Consider the two configurations depicted in Figure 1: in the left-hand configuration the system is not trilaterally centralized, whereas the common overlap between the three spheres in the right-hand picture suggests a strong integration in terms of both bilateral and trilateral relations. In terms of organizing a competitive advantage at the level of a nation or region, central integration related to UIG relations may be considered desirable from a policy perspective (Etzkowitz & Leydesdorff, 2000; Mirowsky & Sent, 2007). In the left-hand configuration, however, a complex dynamics can emerge that leaves the self-organization of the system to mutual adjustments between the partners without the need for a trilateral center of coordination. For example, relations can be asynchronous, but nevertheless fine-tuned. Under these conditions, a differentiated configuration would be able to process more complexity than an integrated one because integration at the center would impose also a (potentially normative) condition.

This mutual information in three dimensions enables us to measure the balance between the dynamics of integration and differentiation at the systems level in terms of the relative frequencies of relations among the partially overlapping sets. In general, mutual information can be considered as an information-theoretical analogue of co-variation. The co-variation between two variations reduces the uncertainty on both sides. Unlike co-variance analysis among three or more variates,[3] information-theoretical measures are dimensionless and allow for comparisons among (quasi-)experimental results which differ in their metric (Garner & McGill, 1956, p. 228).

**Results**

Figures 1 provides the results based on the SCI data over time. The collaboration patterns, as measured by co-authorship relations in the ISI publications, noticeably increased, with some variation, from the mid-1970s to the mid-1990s. However, inter-institutional collaboration in the early 21st century has decreased as a percentage of the total collaboration. These results are interesting as they reveal a major trend in accordance with the influence of the government R&D programs in shaping the expansion and contraction of collaborations between universities, companies, and public research centers during the last four decades. For example, the number of university-authored SCI publications has skyrocketed largely due to the introduction of BK21 programs. University-government collaborations have been progressing more rapidly than university-industry partnerships since the government's focus in university-industry collaborations has been on patenting and not publishing scientific articles.

---

[3] In the case of co-variance analysis, the assumptions (e.g., about the shape of the distribution) are more restricted, and the results more difficult to interpret (Garner & McGill, 1956).

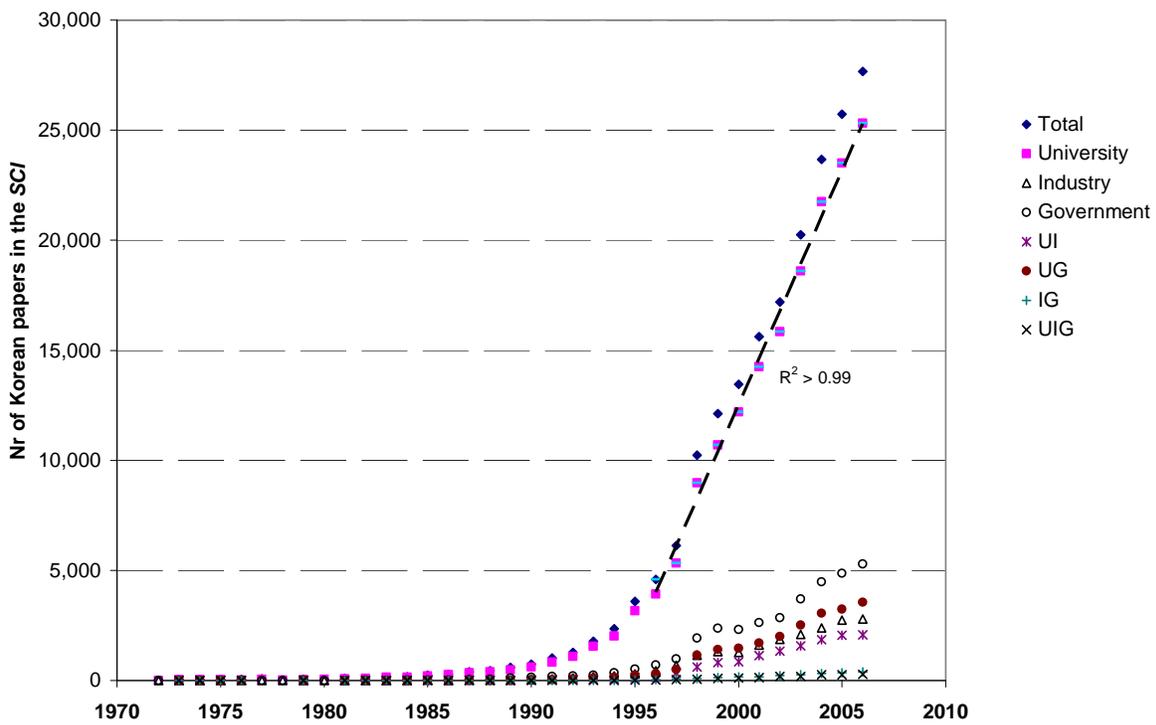

**Figure 1**: Number of papers by Korean authors in the SCI and bilateral and trilateral relations between TH sectors within the economy.

Let us first turn for more detail to mutual information in bilateral relations between the TH sectors as a measure of co-authorship relations in the Korean national system of publications (Figure 2).

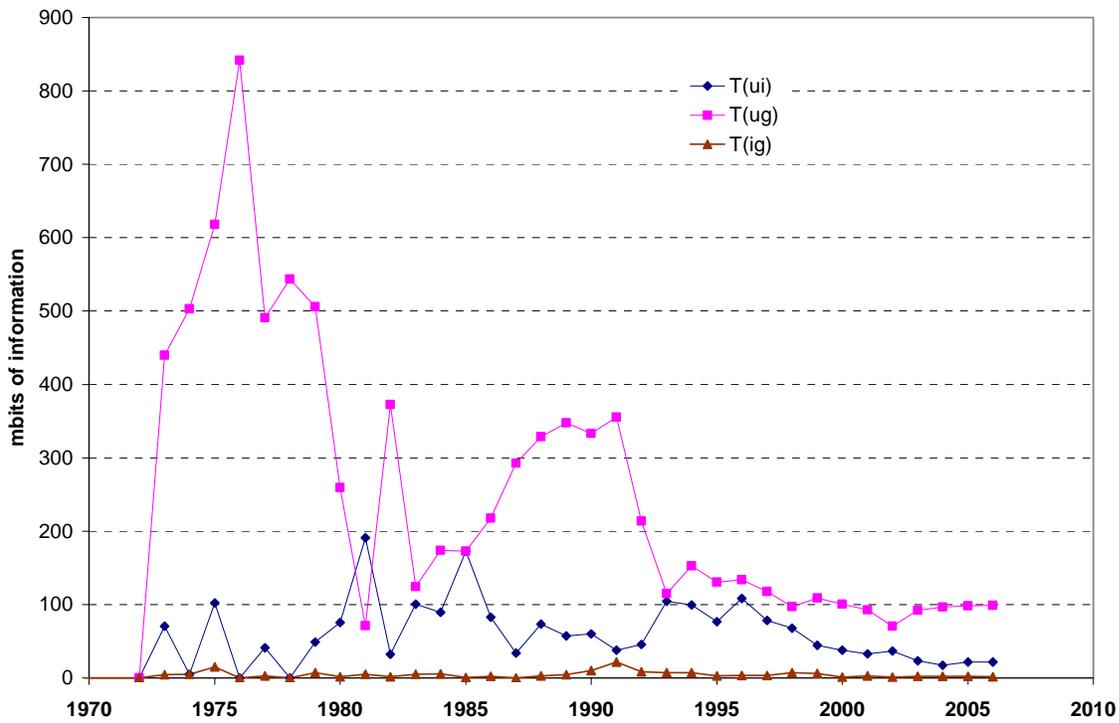

**Figure 2**: Mutual information in bilateral relations between the TH sectors in Korea

Figure 2 shows that university-government research collaboration ($T_{ug}$) has been strongest among the bilateral relations. The transmission between universities and public agencies reached its highest value of 841.85 mbits in 1975, during the period of President Park, Jung-Hee. While $T_{ug}$ values decreased during the administration of President Chun (1980-1987), the UI collaborations began to blossom during this period. The decrease of university-government relations may have been caused by the forced integration of major public institutions with some national universities during this period. The transmission value in UI co-authorship relations increased to $T_{ui}$ = 172.65 mbits in 1985. During this entire period (since 1970), mutual information between industries and government agencies ($T_{ig}$) has been stagnant and scientific cooperation between the academic and industrial publication systems ($T_{ui}$) remained more active than between industry and government ($T_{ig}$). Note that the gap between ($T_{ug}$) and ($T_{ui}$) has decreased since 1995. However, this decreasing difference is largely due to a decrease in university-government collaborations ($T_{ug}$), rather than the expansion of university-industry collaborations ($T_{ui}$).

While the mutual information in the bilateral relations is by definition positive, the TH indicator ($T_{uig}$) as operationalized above can also be negative. This reduction of uncertainty at the systems level can be considered as a synergetic effect of TH relations.

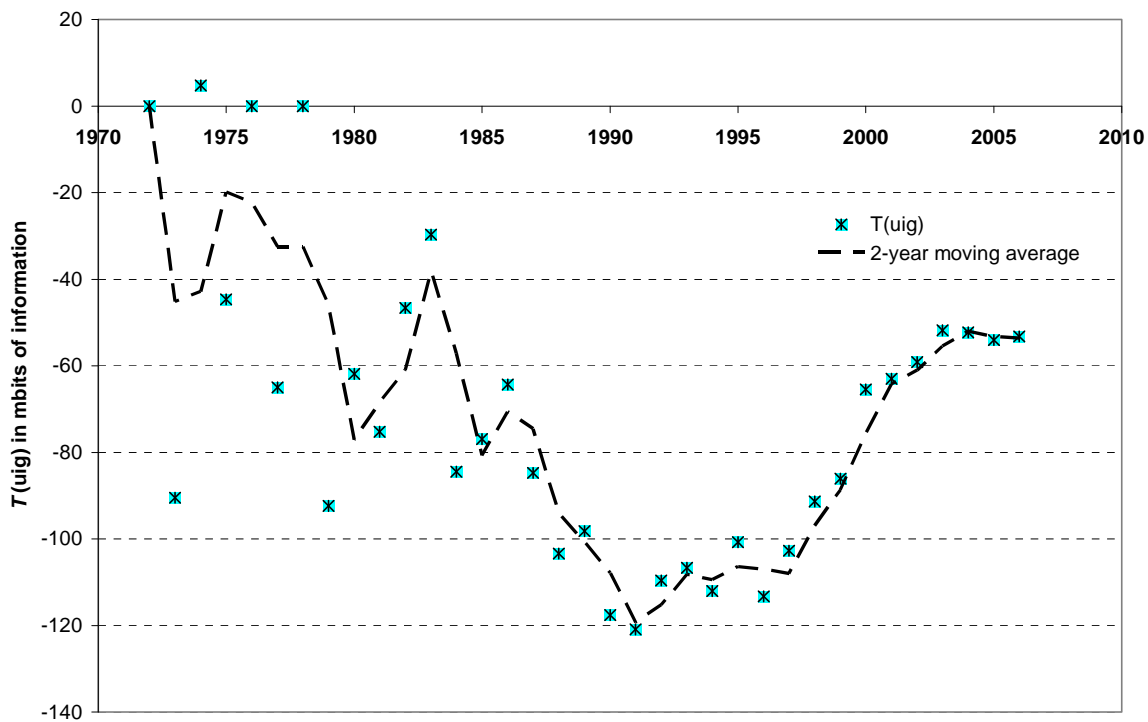

**Figure 3:** Mutual information in trilateral TH relations in Korea.

The longitudinal trend (Figure 3) as expressed using two-year moving averages shows an interesting path among the three institutional spheres. First, there is a reduction of uncertainty among academic, public, and industrial research actors in the Korean publication system from 1970 to 1990. For example, the values of the configurational information $T_{uig}$ during the period of President Roh, Tae-Woo ranged from -98.24 mbits in 1989 to a minimum of -120.91 mbits in 1991. Thus, an intensive and synergistic collaboration among the UIG institutions is indicated in the early 1990s (Figure 3).

Second, the mutual information among the three TH agencies ($T_{uig}$) is relatively stable during the 1990s, but begins to decrease during the last ten years. The trend line shows that the TH dynamics of UIG relations has varied considerably; this variation generally accords with changes in Korean government research policies as explained in an earlier section. Third, the contribution of academic knowledge to the Korean economy has surprisingly declined since 1998 using this indicator. While the TH networks of Korean research institutions were becoming more cohesive (as indicated by the two-year moving averages) in terms of the co-authorship relations in ISI-listed journal publications, a decisive break occurred around 1998. This can be attributed to the Asian financial crisis of 1997 and 1998, and the subsequent instigation of the Brain Korea 21 (BK21) program.

The Korean government launched the BK21 program with the purpose of promoting university research because of the important role universities play in the national innovation process. As noted, this program was successful on its own terms. However, the TH indicator reveals a decrease in the network dynamics during this period. In particular, the bilateral relations between researchers in universities and industries significantly decreased in activity. In our opinion, this was partly due to the evaluation policies of the newly introduced BK21 program that emphasized the number of ISI-listed journal publications for university researchers (Moon & Kim, 2001; RAND, 2007). This recalls the criticisms of Browman and Stergiou (2008) that the use of bibliometric indices in evaluating scholarly performance, without a thorough and insightful understanding of their strengths and weaknesses, can be misleading for the development of a national research policy.

In addition to these policy incentives, universities in Korea became institutionally strict about promoting faculty members during this latter period (Park *et al*., 2008). Several schools required natural and engineering science researchers to publish their research only in SCI journals if they wanted to continue to be employed or gain promotion. One can wonder whether under such institutional-regulatory frameworks, academic researchers will collaborate across institutional borders. The institutional incentives in terms of funding grants and faculty promotion tend to give credit to single-authored publications more than collaborative publications. In evaluations, the contribution of individual researchers in multi-authored papers is often divided by the total number of authors ("fractional counting"). This practice may further discourage interactions among the three institutional spheres of the TH dynamics.

Furthermore, the newly introduced PBS stood in the way of cooperation with other R&D partners for academics in governmental research institutes (GRIs) in this first decade of the 21$^{st}$ century. Under the PBS system, government-affiliated scientists were salaried according to their quantitative performance in terms of the number of acquired profitable projects during the previous one or two years. This program discouraged public-sector researchers from collaborating with researchers in academia or industry because the government-affiliated scientists were increasingly reluctant to share their research activities with others. The institutional incentive was aimed at securing resources in cutting-edge projects for the individual researcher.

Finally, in addition to the fact that the Korean government policy tends to focus on numbers of publications (e.g., BK21) and that this may have an adverse effect on UIG joint research, it seems that other mechanisms may also be at work. For example, institutional R&D partnerships in Korea have undergone a turbulent transition closely connected to the technological content of research output. When Korean universities and public research institutes were producing innovative research in emerging technologies during the 1970s and 1980s, UIG collaborations were popular among academic scientists. In the field of electronics, for example, the GRIs helped foster related R&D in private research institutes and there were active information transfers between public and private sectors through informal consultations and organizational collaborations.

By the mid-1980s, Korean firms had surpassed public R&D capabilities (e.g., in semiconductors) so that UIG collaborations began to wither (Wade, 1990). On the other side of this divide, innovative R&D capabilities in Korean academic and public sectors seem to become strong in terms of scientific performance (e.g., SCI publications), but low in commercialization capabilities (Lee, 2008). The low economic and commercial value of research output generated by Korean academic and government institutions may discourage business researchers from collaborating with their academic partners within Korea (Leydesdorff & Sun, 2009). However, this does not necessary mean that industry researchers are far less enthusiastic in co-inventing commercial products with their academic colleagues in Korea.

## The Social Sciences and the Humanities

Using the same indicator of UIG relations, we examined mutual information among the UIG institutions in the A&HCI and the SSCI. While academics in these 'soft' sciences may be relatively less concerned with scientific collaboration than their colleagues in the 'hard' sciences, the national importance of UIG collaboration is no longer restricted to the natural and engineering sciences. Some scientific programs in the humanities and social sciences rely on multi-researcher endeavors requiring substantial information exchanges among the academic, public, and industrial sectors. For example, in archiving historic sites and digital story-telling, Korea's social scientists and humanities scholars need to collaborate with one another across their institutional boundaries because their research projects are increasingly integrated with e-science practices (Soon & Park, 2009). However, public R&D funding agencies have not been active in helping researchers in the humanities and social sciences expand their research projects beyond their universities by collaborating with partners in government offices and industry (Korea Research Foundation, 2008). The Academic Promotion Plan of the Korea Research Foundation (2009-2013) stresses the need for scholars in the humanities and social scientists in particular to cooperate in the knowledge diffusion and national innovation process (Yang, 2008). It should be noted, though, that this plan extends from 2009-2013 and is thus beyond the publications dates considered in this study.

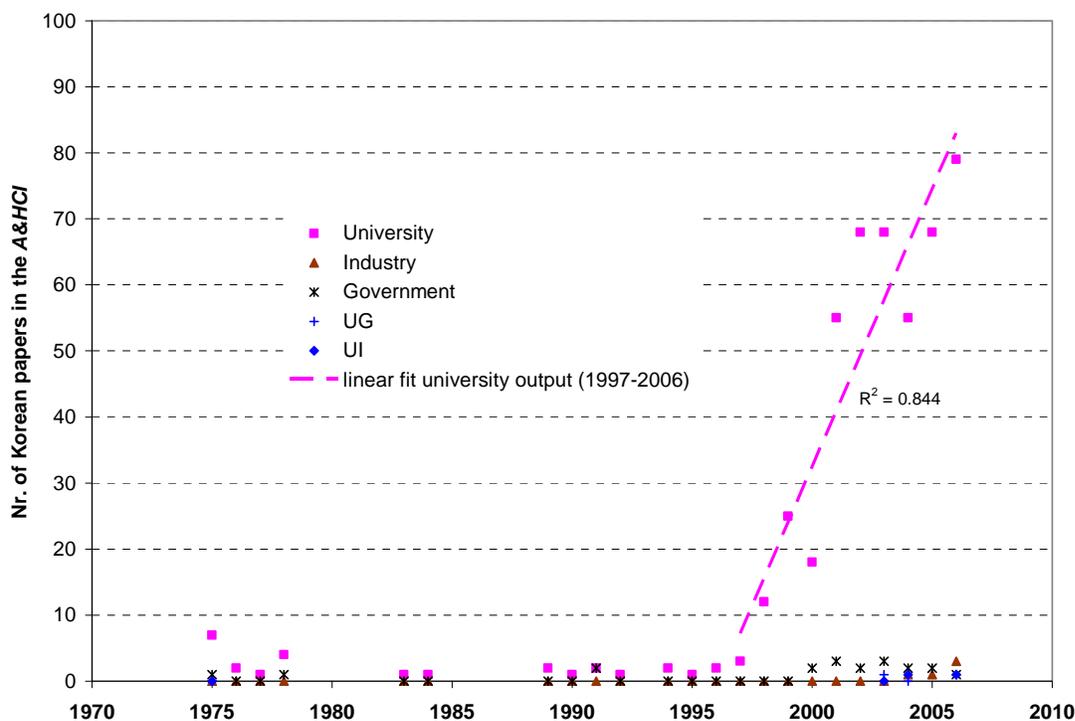

**Figure 4:** Publication patterns in and between TH sectors using the A&HCI.

Figure 4 shows the disappointing result for TH dynamics in the humanities. The TH indicator did not work for the A&HCI data because there are extremely few industrial publications in this database; it consists almost exclusively of academic publications. A&HCI publications written by scholars with university addresses increased rapidly from 1997 to 2006. Collaborative arrangements in the humanities often produce book-style policy reports—mostly in Korean—but not often articles in international journals. The Korean audience (including public officials) in the humanities prefers Korean documents with detailed tables, figure, and explanations. Scholars in this field do not strive to publish their collaborative and often policy-oriented outputs in English[4].

---

[4] One of the limitations in current research is the under-representation of UIG collaboration in the humanities and social sciences in terms of SSCI and A&HCI journal publications.

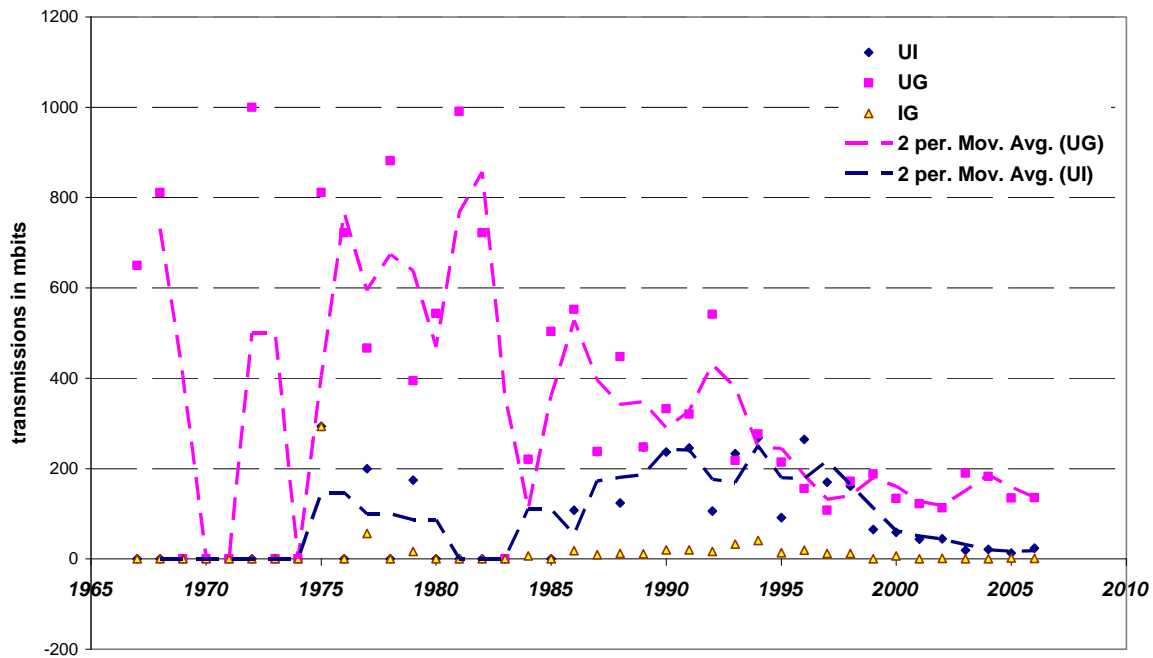

**Figure 5**: Mutual information measured in bilateral relations between TH sectors using the SSCI.

Figures 5 shows the results of the analysis of bilateral relations between TH sectors for the SSCI in terms of mutual information measured in bits of information. Interestingly, UG and UI collaborations show different results from the mid-1980s to the mid-1990s; $T_{ug}$ values were in decline, but $T_{ui}$ values went up during this period. The BK21 program, begun in 1998, could be expected to affect mutual information between university and industrial research outputs adversely, because university researchers funded by BK21 are evaluated on their publication performance rather than on their contributions to transfer information or their social impact. During the most recent decade, we consequentially observe a decrease in the network dynamics of social science research (Figure 6).

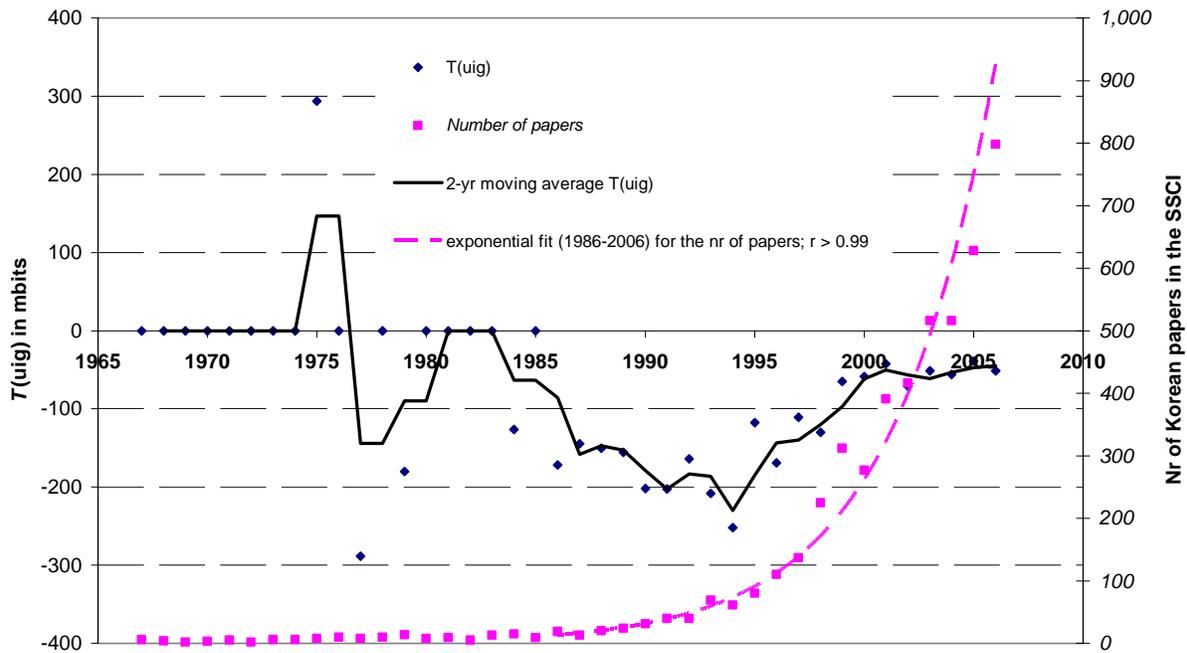

**Figure 6**: Publication rates of Korean papers and synergy effects among TH sectors on the basis of co-authorship relations in the SSCI.

Mutual information in co-authorship relations among the three institutional partners of the TH in the SSCI is shown in Figure 6. The overall pattern of the TH dynamics in the social sciences is more or less similar to the SCI-based indicators, although there is a delay in terms of the longitudinal pattern. In other words, the research practices of Korean social scientists in terms of institutional R&D partnerships have become more like those in the natural and engineering sciences.

For example, national co-authorship relationships became more dynamic among universities, industry and government between the mid-1980s and the mid-1990s. This is shown in Figure 6 as decreasing uncertainty at the systems level. However, there has been less uncertainty reduction as expressed by $T_{\text{uig}}$ values in more recent times. Judging from the number of published papers, the research portfolio in the social sciences has been prolific since 1994. Note that there are also differences between the two datasets; for example, the *exponential* fit for the output measured in the SSCI from 1986 to 2006 is larger than 0.99, whereas a *linear* fit for SCI-based data during the last 10 years exceeds 0.99. This indicates that the social sciences are still catching up with the natural sciences in terms of turning national (and non-ISI) output into output that is appreciated in institutional evaluations.

## Conclusions and Discussion

Nations differ in terms of their research portfolios and their economic structures. Both research fields and industrial structures are internationally organized. This raises a number of control problems at the national level. Government intervention—particularly in small nations such as Korea—can no longer be expected to steer these developments. Governments nevertheless are under pressure to develop programmatic incentives. These incentives are focused on the institutional level; the incentivized systems, however, have a degree of freedom in the international dimension and may show resilience regarding government intervention.

A national system of innovations such as the Korean (or *mutatis mutandis* the Brazilian) case can be considered as a complex construct of integrating and differentiating mechanisms (Etzkowitz & Brisolla, 1999). Differentiation is enforced within each of the helices of the TH model by the rationale of specific subsystems: scholars wish to publish; industries wish to gain financially from collaboration; and policy-makers represent the public interest, but also want to win elections. Korean governments have focused on improving university-industry relations with the purpose of strengthening Korea's system of innovations. However, because of historical developments, the priorities in these programs also have changed.

First, during the military dictatorship of the 1970s the integration of the administration for control purposes was a secondary objective in the relatively new S&T policies. Not amazingly, this led to a strong relative increase of university-government co-authorship relations because university academics could only gain legitimacy and secure funding by aligning themselves with the national policy goals. The liberalization of the 1980s witnessed the withering of these bureaucratically inspired circles and the rise of university-industry collaborations in strategic sectors of the Korean economy. The synergetic effects of bilateral and trilateral UIG relations on one another increased spectacularly during this period.

During the 1990s, the tables were turned. First the demise of the Soviet Union and the opening of China in the decade posed a major challenge to the Korean innovation system. Knowledge and technology-intensive industry became crucial assets for development in a knowledge-based economy. Korean industry, however, had matured during the 1980s and become internationalized. As Leydesdorff and Sun (2009) found in Japan, university-industry relations tend to become more internationally than nationally oriented[5].

The Korean government drew conclusions from these developments after the monetary crisis in Asia in 1997-1998. The new government programs were from then onwards based on differentiation among the three sectors of the research system: academia was stimulated to develop according to international criteria of publications and citations, industry was no longer tied to the national knowledge base, and the project-based system in public-sector research was incentivized to channel research results into commercially viable technologies and innovations. In academia, this policy complemented an ongoing trend to focus on international publishing, and universities reacted by adapting their hiring and promotion policies to these core incentives.

---

[5] We plan to investigate the role of international collaboration in national research system in future research. As a matter of fact, we are in the middle of classifying international authors included in Korea's SCI, SSCI, A&HCI publications.

As a consequence of this differentiation, the integration in TH relations became less central not only to policy making, but also to the dynamics of the Korean innovation system itself. The synergy in TH relations has eroded since 1998. The role of government as interventionist agent providing integration at the national level decreased at the institutional level. Whereas industry naturally tends to react to "the forces of the market" more than to government incentives, academia now also reacts to international standards like Shanghai-rankings, scores of publications and citations in internationally defined databases, etc. The institutional incentives are increasingly adapted to the functionally specific ones and less subservient to national objectives set on programmatic grounds.

Incentives at the institutional level—further reinforced by national programs like the BK21— cannot be applied to some faculties and not to others. Thus, the social sciences and humanities have been drawn into the incentives "rat-race" that is ongoing in the natural and life sciences. These disciplines are intrinsically more oriented to cultural and social contexts on which they reflect and thus contribute analytically, intellectually, and critically. While the pressure to publish may lead in the natural and life sciences to premature knowledge claims (e.g., in stem-cell research), this pressure tends to dissolve the relationship between national cultures and their resources in the humanities and the social sciences. It is easier to publish in prestigious journals using American statistics about the American economy than using a precise analysis of the dynamics of Korean economic development. Since the descriptive statistics are increasingly available on the Internet, one can witness these changes of interest in topical selections in developed smaller nations such as Denmark, the Netherlands, and Korea. Scholarly journals published in Korean are increasingly perceived among academics as playgrounds for PhD students who have not yet reached the level of publishing internationally.

In summary, government policies have influenced the Korean system to an extent which we did not expect given the continuities and resiliencies prevailing in all systems. However, an unintended effect of these policies has been the erosion of national co-authorship relations spanning institutional boundaries. The systems to be incentivized have been internationalized and globalized. Government has been left behind at the national level of integration; they have tried to gain legitimacy by jumping onto emerging research bandwagons. The most recent financial crisis may have turned the tables again, but recent developments fall outside the scope of this study.

The results of this study has shown that the emphasis of the Korean government and universities on publication performance in funding allocations, grants, faculty recruitment, promotion, and tenure discourages collaborations among the three institutional spheres of the TH dynamics. To correct this imbalance, Korea's national research policy should be based less on strict quantitative performance measures and more on a balanced approach between bibliometric indices and the informed judgment of peers with expertise and academic maturity. Universities and government-sponsored research institutes, as suggested by Hwang *et al.* (2004), should enhance their organizational flexibility in order to boost social capital (e.g., motivation, collaboration experiences, and cognitive conditions) by cooperating more with their industrial R&D partners. Lastly, it is recommended that collaborative UIG publications in international journals be encouraged through a reformulation of research evaluation policies in order to stimulate inter-organizational R&D cooperation.

<Acknowledgement>

This paper is a revised version of a paper presented at the 7[th] International Conference of the Triple Helix of University-Industry-Government Relations in Glasgow (UK), 17-19 June 2009. The authors are grateful to Min-Ho So for his assistance in collecting the data.

Trotter, II. R. T., Briody, E. K., Sengir, G. H., & Meerwarth, T. L. (2008). The life cycle of collaborative partnerships: Evolution of structure and roles in industry-university research networks. *Connections*, 28 (1), 40-58.

Ulanowicz, R. E. (1997). *Ecology, The Ascendent Perspective*. New York: Columbia University Press.

Wade, R. (1990). *Governing the Market: Economic Theory and the Role of Government in East Asian Industrialization.* Princeton: Princeton University Press.

Wagner, C. S. (2008). *The New Invisible College*. Washington, DC: Brookings Press.

Yamauchi, I. (1986). Long Range Strategic Planning in Japanese R&D. In C. Freeman (Ed.), *Design, Innovation and Long Cycles in Economic Development* (pp. 169-185). London: Pinter.

Yang, J. M. (2008). Supporting interdisciplinary research projects. *2007 Research Policy*. Seoul: Korea Research Foundation (pp. 130-141). Written in Korean.

Yeung, R. W. (2008). *Information Theory and Network Coding,* New York, NY: Springer. Retrieved from http://iest2.ie.cuhk.edu.hk/~whyeung/post/main2.pdf.
40